\documentclass{PoS}
\usepackage{amsmath}
\newcommand{\be}{\begin{equation}}
\newcommand{\ee}{\end{equation}}
\newcommand{\bea}{\begin{eqnarray}}
\newcommand{\eea}{\end{eqnarray}}

\title{Lattice Supersymmetry: Some Ideas from Low Dimensional Models}

\ShortTitle{Lattice Supersymmetry: Some Ideas from Low Dimensional Models}

\author{\speaker{Alessandro D'Adda}\\
        Dipartimento di Fisica Teorica, Universita' di Torino and INFN Sezione di Torino\\
         I-10125 Torino, Italy. \\
        E-mail: \email{dadda@to.infn.it}}
\author{Alessandra Feo\\
         Dipartimento di Fisica Teorica, Universita' di Torino and INFN Sezione di Torino\\
         I-10125 Torino, Italy.\\
        E-mail: \email{feo@to.infn.it}}
\author{Issaku Kanamori\\
         Dipartimento di Fisica Teorica, Universita' di Torino and INFN Sezione di Torino\\
         I-10125 Torino, Italy.\\
        E-mail: \email{kanamori@to.infn.it}}
\author{Noboru Kawamoto \\
        Department of Physics, Hokkaido University\\
        Sapporo, 060-0810 Japan \\
        E-mail: \email{kawamoto@particle.sci.hokudai.ac.jp}}
\author{Jun Saito\\
         Department of Physics, Hokkaido University\\
        Sapporo, 060-0810 Japan\\
        E-mail: \email{saito@particle.sci.hokudai.ac.jp}}

\abstract{In the framework of the so called link approach we study
exact lattice supersymmetry for the simplest supersymmetric model:
$N=1$ supersymmetry in $D=1$. The model is described by a lattice
with spacing $\frac{a}{2}$, thus containing twice as many sites as
the conventional one. The boson and fermion are related through a
$2\pi/a$ momentum shift, which can provide an interpretation of
them being species doublers to each other.  An exactly
supersymmetric lattice action can be written within this scheme in
momentum representation, which  however turns out to be non local
in coordinate space. }

\FullConference{The XXVII International Symposium on Lattice Field Theory - LAT2009\\
         July 26-31 2009\\
         Peking University, Beijing, China}

\begin{document}

\section{Introduction}

In recent years a number of lattice models with at least one exact
supersymmetry at non-zero lattice spacing have been proposed \footnote{
For a review see \cite{Catterall:2009it} and references therein. }.
Topological twist and orbifold projection are the main
new ideas behind these approaches, which however allow in general
only one exact lattice supersymmetry. The link approach was
proposed to overcome this limitation and realize exactly on the
lattice all supersymmetries in some models with extended
supersymmetry \cite{D'Adda:2004jb,D'Adda:2005zk,D'Adda:2007ax,Arianos:2008ai}.
The key new ingredient of this approach with
respect to the ones mentioned above (to which it is however deeply
related \cite{Damgaard:2007eh}) is the introduction of extended lattices with additional
``fermionic'' links on which supersymmetry charges sit. As a
consequence of the link nature of the (super)symmetry charges,
modified Leibnitz rules have to be applied when (super)charges act
on a product of (super)fields. Hence both the consistency and the
relevance of the link approach to the description of exact lattice
supersymmetry has been questioned \cite{Bruckmann:2006ub,Bruckmann:2006kb}.
However it has been shown
recently by some of the present authors that a consistent
mathematical set up for the link approach can be  given in terms
of Hopf algebras~\cite{D'Adda:2009kj}.
The introduction of new links and new sites
implies that a larger number of degrees of freedom  is present in
the link approach formulation, namely that the theory contains
doublers both for bosons and fermions unless some mechanism is
found to get rid of them. This is one motivation for looking at
the simplest supersymmetric system: an $N=1$ supersymmetry in one
space-time dimension. In spite of its simplicity the investigation of this model
in the framework of the link approach offers some insight into
some relevant problems, including the doublers mentioned above,
and it is worth pursuing. This will be the subject of the present
talk.

\section{D=1, N=1 model}

The simplest supersymmetric model is a one dimensional model with just one supersymmetric charge.
It is described in terms of a superfield:
\be \Phi(x,\theta)= \varphi(x) + i \theta \psi(x) \label{superfield}\ee
with a supersymmetry charge given by:
 \be Q = \frac{\partial}{\partial \theta} +
i \theta \frac{\partial}{\partial x} \label{susycharge}\ee
and
\be Q^2 = i \frac{\partial}{\partial x}. \label{qsquare}\ee
The free action is given by \be S = \frac{i}{2} \int dx\, d\theta ~
D \Phi~ \frac{\partial \Phi}{\partial x} \label{action} \ee where
$D$ is the super derivative.
Because of the fermionic nature of the superspace integration volume no potential can be written in terms
of the superfield and the theory is essentially free.

On the lattice the derivative $\frac{\partial}{\partial x}$ is replaced by finite shift $\Delta$, defined by:
\be
\Delta \Phi(x) = \Phi(x+a)\Delta, \label{findiff}
\ee
where $x$ and $x+a$ are two neighboring sites separated by the lattice spacing $a$.

In the link approach a shift $a_Q$ is associated to the
supersymmetry charge $Q$, and since eq. (\ref{qsquare}) is
replaced by \be Q^2 = i \Delta  \label{qlat} \ee we have $a_Q
=\frac{a}{2}$. The supersymmetric extended lattice is then given
by integer multiples of $\frac{a}{2}$, thus doubling the number of
the the original lattice sites. We assume here that $Q$ acts on
the fields as a shift\footnote{This is in agreement with the
matrix representation of lattice superspace given in
\cite{Arianos:2008ai}. Notice in particular that in this way $Q$
does not contain any Grassmann odd parameter.} of $\frac{a}{2}$,
just as $\Delta$ acts as a shift of $a$ . The question now is:
What makes this lattice different from just an ordinary lattice
with spacing $\frac{a}{2}$? The answer is in the different
behaviour of bosonic and fermionic fields with respect to shifts
of $\frac{a}{2}$, namely with respect to supersymmetry
transformations. In the continuum a constant bosonic field
commutes with $Q$, hence for a constant field we have on the
lattice: \be \varphi(x+\frac{a}{2}) - \varphi(x) = 0 \label{costb}
\ee which implies that $\varphi$ is constant on the lattice. A
constant fermionic field instead anticommutes with $Q$: $\{
Q,\psi\} = 0$ . If we assume that on the lattice $Q$ simply acts
as a shift of $\frac{a}{2}$ , then $\{ Q,\psi\} = 0$  implies for
a constant fermionic field $\psi_l(x)$:
\be \psi_l(x+\frac{a}{2})
+ \psi_l(x) = 0 \label{constf},
\ee
namely
\be
\psi_l(x) =
(-1)^{\frac{2 x}{a}} \psi_0\, , \ee where $\psi_0$ is a constant  and
$\frac{2 x}{a}$ is an integer on the lattice.

Physical fields are fluctuations around constant configurations.
One can then tentatively write a superfield on the lattice as \be
\Phi(x) = \varphi(x) + \frac{a^\frac{1}{2}}{2}
(-1)^\frac{2x}{a}
\psi(x) \label{lattsup} \ee where $\varphi(x)$ and $\psi(x)$ are
smooth fields, in the sense that for instance
$\psi(x+\frac{a}{2})-\psi(x)$ is of order $a$ in the continuum
limit. The smooth field $\psi(x)$ is related to the original
lattice fermion $\psi_l(x)$ by the relation \be \psi_l(x)
=(-1)^{\frac{2 x}{a}} \psi(x) \label{psil} \ee so that $\psi_l(x)$
satisfies the smoothness condition
$\psi_l(x+\frac{a}{2})+\psi_l(x) = O(a)$. Eq. (\ref{lattsup})
 resembles the usual superfield expansion with the sign
factor $(-1)^{2x/a}$ playing the role of $\theta$.

The supersymmetry  transformations are given in the continuum by:
\be
\delta \Phi(x,\theta) = \alpha [Q,\Phi] \,\label{sustc} .
\ee
The lattice equivalent is
\be
\delta \Phi(x) = a^{-\frac{1}{2}} \alpha (-1)^\frac{2x}{a}
\left( \Phi(x+\frac{a}{2}) - \Phi(x)\right), \label{suslatt}
\ee
where the constant fermionic parameter $\alpha$ in the continuum
 has been consistently
 replaced on the lattice by $\alpha (-1)^\frac{2x}{a} $ .
From here one can read the supersymmetric transformations for the
 component fields as
 \bea &&\delta \varphi(x) = -\frac{\alpha}{2}
\bigg[ \psi(x+\frac{a}{2}) + \psi(x) \bigg] \xrightarrow[a\to 0]{}
-\alpha \psi(x) \, ,\label{suslatt1} \\
&& \delta \psi(x) = 2 a^{-1} \alpha \bigg[ \varphi(x+\frac{a}{2})
- \varphi(x)  \bigg]  \xrightarrow[a\to 0]{}
\alpha \frac{\partial
\varphi(x)}{\partial x} \, . \label{suslatt2} \eea

The supersymmetry transformations (\ref{suslatt1}) and
(\ref{suslatt2}) have the correct structure, but they are still
not the right ones. In fact the variation of $\varphi(x)$ at the
l.h.s. of (\ref{suslatt1}) is not real: an $i$ factor is missing.
In order to restore the hermiticity of the supersymmetry
transformations symmetric finite differences must be used,
introducing a shift of $\frac{a}{4}$ of the fermionic fields sites
with respect to the bosonic ones. Hence, instead of writing the
superfield on the lattice as in (\ref{lattsup}) we shall introduce
$\Phi(x)$, with $x=n \frac{a}{4}$, defined by:
 \be \Phi(x) = \left\{ \begin{array}{lc} & \varphi(x)~~~~~~~~\textrm{for}~~~
 x=n a/2 \\&\frac{1}{2} a^{1/2} e^{\frac{2i \pi x}{a}}
 \psi(x) ~~~\textrm{for}~~~x=(2n+1)a/4 .
 \end{array} \right. \label{spf} \ee
 Again the supersymmetry transformations can be written in terms
 of $\Phi(x)$:
 \be
 \delta\Phi(x) = \alpha a^{-1/2} e^{\frac{2 i \pi x}{a}} \left[
 \Phi(x+a/4) - \Phi(x-a/4) \right]. \label{st}
 \ee
 By separating $\Phi(x)$ into its component fields according to
 (\ref{spf}) we find:
 \bea
&&\delta \varphi(x) = \frac{i \alpha}{2} \bigg[
\psi(x+\frac{a}{4}) + \psi(x-\frac{a}{4}) \bigg] \xrightarrow[a\to 0]{}
i \alpha \psi(x) \, ,\label{suslattf1} \\
&& \delta \psi(x) = 2 a^{-1} \alpha \bigg[ \varphi(x+\frac{a}{4})
- \varphi(x-\frac{a}{4})  \bigg]  \xrightarrow[a\to 0]{}
\alpha
\frac{\partial \varphi(x)}{\partial x} \, , \label{suslattf2} \eea
where $x$ is an even multiple of $a/4$ in (\ref{suslattf1}) and an
odd one in (\ref{suslattf2}). As in the continuum case the
commutator of two SUSY transformation is a translation, namely, on
the lattice, a finite difference of spacing $a$. For instance we
have for $\varphi(x)$ (the same applies to $\psi(x)$): \be \delta_{\beta}\delta_{\alpha}\varphi(x) -
\delta_{\alpha}\delta_{\beta}\varphi(x) = 2 i \alpha \beta \left[
\varphi(x+a/2)-\varphi(x-a/2) \right]. \label{trs} \ee 
To summarize: even in this extremely simple case exact
supersymmetry on the lattice requires the doubling of lattice
sites for both bosons and fermions, with the lattice spacing
halved from $a$ to $a/2$, the alternating sign structure for the
fermion fields, and, to preserve hermiticity, a relative shift of
$a/4$ of the boson and fermion lattice sites so that ultimately
the effective lattice spacing is $a/4$. The price we had to pay
for introducing supersymmetry is the doubling of both boson and
fermion degrees of freedom. How to reduce them to the original
number without spoiling supersymmetry is the next task. For this
purpose we shall move from coordinate to momentum representation.

\section{Momentum Space}

Let us consider first the Fourier transform of the component
fields $\psi(x) $ and $\varphi(x)$, and denote them by $\tilde
\psi(p)$ and $\tilde \varphi(p)$ respectively. The lattice spacing
being $a/2$, the Brillouin zone extends over a $\frac{4\pi}{a}$
interval and besides the physical states at $p=0$ will include
doublers at $p=\frac{2 \pi}{a}$. Hence we have:
\be \tilde
\varphi(p+\frac{4\pi}{a})=\tilde \varphi(p),\,\,\,\,\,\,\,\tilde
\psi(p+\frac{4\pi}{a})=-\tilde \psi(p), \label{period} \ee
where
the minus sign in the case of $\tilde \psi$ is due to the $a/4$
shift in coordinate space. The supersymmetry transformations
(\ref{suslattf1}) and (\ref{suslattf2}) are then given by:
\bea
&& \delta \tilde \varphi(p) = i \cos \frac{a p}{4} \, \alpha \tilde \psi(p)\label{susymom1}, \\
&& \delta \tilde \psi(p) = -i\frac{4}{a} \sin\frac{a p }{4}\,
\alpha \tilde \varphi(p).  \label{susymom2}
\eea
Eqs. (\ref{susymom1}) and (\ref{susymom2}) are consistent with both the periodicity conditions
(\ref{period}) and with the reality conditions expressed in momentum space by:
$ \tilde \varphi(p)^\dag = \tilde \varphi(-p)$ and
$\tilde\psi(p)^\dag =\tilde\psi(-p)$.

Let us consider now the Fourier transform $\tilde\Phi(p)$ of the
superfield $\Phi(x)$ given in (\ref{spf}) . At each value of $p$,
$\tilde\Phi(p)$ has a bosonic component $\tilde\Phi_b(p)$ and a
fermionic one $\tilde\Phi_f(p)$ given by: \be \tilde\Phi_b(p) =
\tilde\varphi(p), \,\,\,\,\,\,\,\,\,
\tilde\Phi_f(p)=\frac{a^{1/2}}{2} \tilde \psi(p+\frac{2 \pi}{a}).
\label{Phicomp} \ee The periodicity in $p$ of $\tilde\Phi_b(p)$
and $\tilde\Phi_f(p)$ are the same as $\tilde\varphi(p)$ and
$\tilde\psi(p)$ respectively. The supersymmetry transformations
can be easily written in terms of $\tilde\Phi(p)$: \be
\delta\tilde\Phi(p) = -2 i \alpha a^{-1/2} \cos\frac{a p}{4}
\tilde\Phi(p+\frac{2 \pi}{a}) \label{susyspf} \ee which is
equivalent to (\ref{susymom1}) and (\ref{susymom2}). The physical
fields are fluctuations around $p=0$ of the bosonic component
$\tilde\Phi_b(p)$, and around $p=-\frac{2\pi}{a}$ for the
fermionic component $\tilde\Phi_f(p)$. In terms of $\tilde{\Phi}$,
these two physical degrees of freedom have a natural
interpretation of being species doublers to each other.
 The configurations at $p=0$
and $p=-\frac{2\pi}{a}$ correspond respectively to constant and
alternating sign configurations on the lattice of spacing
$\frac{a}{2}$, as discussed above. Large fluctuations are,
however, allowed on the lattice with the result of doubling the
number of degrees of freedom with respect to the original lattice
of spacing $a$. In particular doublers at $p=0$ and
$p=-\frac{2\pi}{a}$ with the ``wrong'' statistics will appear. In
order to reduce the degrees of freedom to the original number it
appears most natural at this point to introduce a cutoff on the
momentum, limiting the bosonic modes to the standard Brillouin zone
$(-\frac{\pi}{a},\frac{\pi}{a})$ and the fermionic ones to
$(-\frac{3\pi}{a},-\frac{\pi}{a})$. In other words wavelengths
shorter that $a$ will correspond to fermionic degrees of freedom,
wavelengths longer than $a$ to bosonic degrees of freedom. This
amounts to impose the constraints \be
\tilde\Phi_b(p)=0~~~~~~p\in\{-\frac{3\pi}{a},-\frac{\pi}{a}\}
~~~~~~\textrm{and}~~~~~~\tilde\Phi_f(p)=0~~~~~~p\in\{-\frac{\pi}{a},\frac{\pi}{a}\}.
\label{constr} \ee These constraints are local in momentum space,
hence highly non local in coordinate space and they allow to
express the value of the fields in the half-integer multiples of
$a$ in terms of the values in the integer multiples. For example,
from the first of (\ref{constr}): \be \varphi(ma + \frac{a}{2}) =
\frac{1}{2 \pi} \sum_n \frac{(-1)^{n-m}}{m-n+\frac{1}{2}}
\varphi(na). \label{int}\ee Non locality in this case does not
arise, as in the SLAC derivative, from the definition of the
derivative on the lattice, but rather from the definition of the
supersymmetric covariant derivative, which involve finite
differences over a $\frac{a}{2}$ spacing.

\section{The Action}

As remarked at the beginning the only possible superfield action
for this system is the free one given in (\ref{action}). This can
be put on the lattice in superfield notation as:
 \be S =\frac{a}{2\pi}
\int^{\frac{ \pi}{a}}_{-\frac{3\pi}{a}} dp~ \tilde \Phi(-p)
\sin\frac{a p}{4}\, \sin\frac{a p}{2}\, \tilde
\Phi(p).\label{latact} \ee The different factors in (\ref{latact})
are in one to one correspondence with the terms in (\ref{action}):
$\sin\frac{a p}{4}$ is a finite difference of spacing
$\frac{a}{2}$ and corresponds to the super derivative $D$ while
$\sin\frac{a p }{2}$ is the lattice version of the normal
derivative $\frac{\partial}{\partial x} $. The superfield
$\tilde\Phi(p)$ is the one given in (\ref{Phicomp}) and the
constraints (\ref{constr}) are understood. For each value of $p$,
$\tilde\Phi(p)$ is purely bosonic or fermionic. The integration
region covers the bosonic range $\{-\frac{\pi}{a},\frac{\pi}{a}\}$
and the fermionic range $\{-\frac{3\pi}{a},-\frac{\pi}{a}\}$.   The
action changes sign when $p \to p + \frac{4 \pi}{a}$, so an
integration over the whole $\frac{8\pi}{a}$ would identically
vanish. It can be checked directly that the action $\ref{latact})$
is invariant under the supersymmetry transformations
(\ref{susyspf}). In terms of component fields (\ref{latact}) can
be written as:
 \be S =\frac{a}{2\pi} \int^{\frac{\pi}{a}}_{-\frac{\pi}{a}} dp \bigg[ \tilde
\varphi(-p) \sin\frac{a p}{4}\, \sin\frac{a p}{2} \, \tilde
\varphi(p) - \frac{a}{4} \tilde \psi(-p) \cos\frac{a p}{4}
\sin\frac{a p}{2} \tilde \psi(p) \bigg] \,  \label{latact2} \ee
and it is invariant under SUSY transformations (\ref{susymom1})
and (\ref{susymom2}). The fields in (\ref{latact2}) are defined in
the $\{-\frac{\pi}{a},\frac{\pi}{a}\}$ range of the momentum and
can be associated in the coordinate space to fields defined on a
lattice of spacing $a$\footnote{Notice however that if  the
momentum integration in (\ref{latact2}) was extended to the whole
Brillouin zone of the extended lattice, that is the interval
$\{-\frac{3\pi}{a},\frac{\pi}{a}\}$, the result would  be
identical to the one given in (\ref{latact2}). In fact the
doublers at $p=\frac{2\pi}{a}$ have the ``wrong'' statistics with
respect to the symmetry of the lagrangian density, and the
integral over the interval $\{-\frac{3\pi}{a},-\frac{\pi}{a}\}$
vanishes identically. The constraints (\ref{constr}) may then be
regarded as superfluous: bosonic modes with wavelength less than
$a$ and fermionic modes with wavelength longer than $a$ naturally
decouple.}. Let us denote them $\hat{\varphi}(an)$ and
$\hat{\psi}(am)$ to distinguish them from $\varphi(x)$ and
$\psi(x)$ defined on a lattice of spacing $\frac{a}{2}$. A simple
Fourier transform allows then to write the action
(\ref{latact2}) in the coordinate space, exhibiting its non local
nature:
 \bea
 S
 &=& \sum_{n,m}\frac{2\sqrt{2}}{2\pi}
     \Biggl[
      \hat{\varphi}(am)\hat{\varphi}(an)(-1)^{m-n}
      \left( -\frac{3}{16(m-n)^2-9} + \frac{1}{16(m-n)^2-1}\right)
      \nonumber\\[-.2em]
 && \qquad\qquad
    -i \hat{\psi}(am)\hat{\psi}(an)(-)^{m-n} a(m-n)
      \left( \frac{1}{16(m-n)^2-9} + \frac{1}{16(m-n)^2-1}\right)
\Biggr]. \label{latactc} \eea

\section{Some Conclusions}
This simple one dimensional model suggests that, within the
extended lattice of the link approach, component fields of a
superfield expansion are associated to different regions of the
Brillouin zone as if they were species doublers of each other.
Momentum representation has a privileged role in this
approach\footnote{Momentum representation has been used in
studying supersymmetric theories in low dimension. See for
instance \cite{Hanada:2007ti} and, for some recent developments
\cite{Kadoh:2009sp,Bergner:2009vg}.}, and an exact supersymmetric
action can easily be constructed in this representation.  The
price to be paid is some non locality in the definition of the
supersymmetric transformations and ultimately of the action when
the coordinate representation is used.  This model is very simple,
in the sense that it contains only one supersymmetric charge with
no interaction. So the extension to higher dimensions, or at least
to extended supersymmetries in one dimension is essential.

This will probably require  a non commutative lattice. the
argument is the following: consider a $D=2 $, $N=1$ supersymmetry
with supersymmetry algebra  $Q^2_1= \frac{\partial}{\partial x}$,
$Q^2_2= \frac{\partial}{\partial y}$ and $\{Q_1,Q_2\}=0$. Then a
superfield expansion on the lattice  in the spirit of
(\ref{lattsup}) would be: \be \Phi(x,y) = \varphi(x,y) +
(-1)^{\frac{2 x}{a}} \psi_1(x,y) + (-1)^{\frac{2 y}{a}}
\psi_2(x,y) +
 (-1)^{\frac{2 x}{a}}  (-1)^{\frac{2 y}{a}} F(x,y).
\ee For this superfield to generate a consistent supersymmetric
interaction term  the sign factors $ (-1)^{\frac{2 x}{a}}$ and $
(-1)^{\frac{2 y}{a}}$ have to anticommute (just as $\theta_1$ and
$\theta_2$ would). This would imply non-commutative space-time on
the lattice, giving \be [x,y] = \frac{i}{4 \pi} a^2 \ee with
commutativity recovered  in the continuum limit $a \to 0$. The
relevance of non commutative lattices in the link approach has
already been considered in \cite{Arianos:2007nv}.

\acknowledgments{
This work is supported in part by
Japanese Ministry of Education,
Science, Sports and Culture
under
the grant No.~18540245 and INFN
research fund.
I.K. is supported by the Nishina Memorial Foundation.
}

\end{document}